# Cold ion beam in a storage ring as a platform for large-scale quantum computers and simulators: challenges and directions for research and development

T. Shaftan and B. Blinov

*Abstract*

The purpose of this paper is to evaluate the possibility of constructing a large-scale storage-ring-type ion-trap system capable of storing, cooling, and controlling a large number of ions as a platform for scalable quantum computing (QC) and quantum simulations (QS). In such a trap, the ions form a crystalline beam moving along a circular path with a constant velocity determined by the frequency and intensity of the cooling lasers. In this paper we consider a large leap forward in terms of the number of qubits, from fewer than 100 available in state-of-the-art linear ion-trap devices today to an order of $10^5$ crystallized ions in the storage-ring setup. This new trap design unifies two different concepts: the storage rings of charged particles and the linear ion traps used for QC and mass spectrometry. In this paper we use the language of particle accelerators to discuss the ion state and dynamics. We outline the differences between the above concepts, analyze challenges of the large ring with a revolving beam of ions, and propose goals for the research and development required to enable future quantum computers with 1000 times more qubits than available today. The challenge of creating such a large-scale quantum system while maintaining the necessary coherence of the qubits and the high fidelity of quantum logic operations is significant. Performing analog quantum simulations may be an achievable initial goal for such a device. Quantum simulations of complex quantum systems will move forward both the fundamental science and the applied research. Nuclear and particle physics, many-body quantum systems, lattice gauge theories, and nuclear structure calculations are just a few examples in which a large-scale quantum simulation system would be a very powerful tool to move forward our understanding of nature.

1. Introduction

Quantum computers harness the power of quantum mechanics to perform a variety of important computational tasks significantly more efficiently than their classical counterparts [1]. Quantum simulators, on the other hand, can solve difficult problems in physics, chemistry, and biology exponentially faster than the most advanced classical supercomputers [2]. Both types of devices, the quantum computers and the quantum simulators, are similar in that they use quantum bits, or qubits, to encode information and perform calculations.

A physical qubit is a two-level quantum system that encodes information in its quantum state in a way that is similar to a classical bit, but in a much larger space: the Hilbert space, which for N qubits has $2^N$ dimensions. This allows quantum devices to process vast amounts of information in parallel and to complete important computational tasks much faster than their classical counterparts [3].

Various physical systems are being actively investigated as potential qubit candidates, including single atoms [4] and ions [5], superconducting circuits [6], semiconductor devices [7], linear optical systems [8], nitrogen vacancies in diamonds [9], and others. The performance of these disparate systems varies widely in terms of the qubit state coherence, operational time scales, quantum gate fidelity, and scalability. Among these, superconducting qubits and trapped ions have emerged as leaders, due to their high operational fidelities and long coherence times. Current state-of-the-art superconducting and ion-trap systems operate using relatively small numbers of physical qubits — fewer than 100. Scaling up this number to hundreds of thousands of physical qubits, where running of practical and useful QC algorithms with fully implemented error correction and fault-tolerant operations may become possible, is an outstanding challenge in the field of quantum computing.

## *2. Ion Traps and Storage Rings*

One of the most promising and advanced approaches in quantum computing is based on ion traps [10, 11]. Ion traps are small electromagnetic devices that combine radiofrequency (RF) and static (DC) electric fields to produce conditions for charged particles to be confined. Laser cooling [12] is used to lower the trapped ions' kinetic energy so that the ions "crystallize" to form the so-called Coulomb crystals [13]. In a linear RF ion trap [14], the confining potential is elongated and the ions form a chain. The qubit is spanned by a pair of stable or metastable internal states in an individual ion. The qubit can be formed by the Zeeman [15] or hyperfine [16] sublevels in the ground state of the ion or by the combination of a ground state sublevel and a metastable excited state [17].

The motion of a single ion in an RF trap is accurately approximated by three-dimensional simple harmonic oscillator with secular frequencies $\omega_x$, $\omega_y$ and $\omega_z$, defined by the trap geometry, the ion's mass, and the driving field frequency (the typical range is $2\pi \times 10 - 100$ MHz) and amplitude (typically of order 1000 V), resulting in secular frequencies in the range of $2\pi \times 0.1 - 10$ MHz. With *N* ions in the trap, there are 3*N* normal modes of harmonic oscillator motion.

To achieve high-fidelity qubit gates with trapped ions, the latter must be laser-cooled to very low temperatures, where the quantization of the ions' motional states in the harmonic oscillator potential becomes important. Specifically, the Lamb-Dicke limit (LD limit) [18] typically must be reached, in which the atom-laser interaction can only cause transitions that change the motional quantum state by 0 and $\pm 1$, and all other transitions are strongly suppressed. The Lamb-Dicke parameter η, which can be expressed as a ratio of the ion's kinetic energy change due to absorption of a single photon to the energy quantum of the harmonic oscillator (in other words, the phonon energy), quantifies the coupling between the internal state transition and the motional states of an ion. If the condition η<<1 holds true, the single photon recoil energy does not change the motional state of the ion, which is a requirement for high-fidelity quantum gates.

Ion cooling into the LD limit is typically accomplished by Doppler cooling [19], followed by several stages of sideband cooling [18], although the second step may be replaced by Sisyphus cooling [20] or electromagnetically induced transparency (EIT) cooling [21]. For trapped particles, a single laser beam is sufficient for efficient Doppler cooling, provided that the k-vector of the Doppler cooling laser beam has significant components along all three (x, y, and z) principal axes of the trap. The sideband cooling typically requires one or two laser beams (in the case of Raman sideband cooling [18]), while Sisyphus and EIT cooling both require two laser beams.

Historically, the development of linear ion traps used in contemporary QC stems from RF quadrupole mass spectroscopy devices [22]. The latter borderline with particle storage rings, with the subtle difference of keeping only a few ions frozen in place by intense laser cooling. It is insightful to look at the origin of both ion traps and ion rings, which, in a large part, comes from the pioneering work of W. Paul. In the 50s he had conceived the idea of the quadrupole mass-spectrometer [23, 24], which turned into a powerful device known today as the Paul trap. About a decade later he proposed a modification of the same spectrometer, titled "Plasmabetatron" [25] according to the ring-like shape and superimposed electric and magnetic fields, which evolved into the PALLAS RFQ ring [26, 27] in the 90s. Indeed, both concepts, cross-compared in our paper, stem from the ion mass spectrometry, however, the ion traps evolved far towards housing low-temperature ion crystals, suitable for carrying out quantum computations. One objective of our paper is to show that the other evolutionary branch of mass-spectrometry, i.e. ion rings, holds the promise of being used as a quantum computation device with a large capacity of qubits [28, 29, 30]. An alternating gradient synchrotron or an RFQ ring, as a rule, are both of substantially larger size than the typical ion trap and, therefore, are capable of storing much larger numbers of charged particles, with up to $10^{10}$ particles orbiting on a ring-like trajectory and being bunched by an RF system [31]. When cooled to a sufficiently low temperature, where the mutual repulsion energy is greater than the ions' motional energy, an ion beam becomes "crystallized" and the crystal's type is determined by the combination of the density of ions and the strength of machine focusing. As the ion density increases, ions form: first, a one-dimensional chain, then a two-dimensional zig-zag, and finally a three-dimensional "shell" structure. Ions in a crystal interact with each other through Coulomb repulsion, which, in terms of the normal modes of motion, can be described as phonons in various frequencies and degrees of freedom [32-35].

Doppler laser cooling in a storage ring may be realized by two counterpropagating laser beams. Scanning the laser frequency, one can both probe and cool the ion beam while recording fluorescence with a photomultiplier in photon counting mode or using a single-photon counting camera. The techniques developed in experiments with crystalline beams enable injection, storage, acceleration, cooling, and control over hundreds of thousands of ions in a storage ring.

One of the issues limiting the scaling up of the trapped ion quantum computers (and, for that matter, all quantum computers) is the decoherence of qubit states, which sets a limit for the length of the working cycle of the computers. Quantum error correction (QEC) techniques include developing redundant chains of qubits with correction algorithms. However, in the present-day NISQ (Noisy Intermediate Scale Quantum) devices, QEC cannot yet be implemented. With the significantly larger number of qubits in a storage-ring device, it may be feasible to demonstrate a fully fault-tolerant QC with QEC.

Several proposals already exist for producing large-scale QC devices based on trapped ions. One such proposal, called a Quantum Charge-Coupled Device (QCCD) [36], is based on interacting ion traps, where multiple set of qubits are stored and moved around the (very complex) trap structure to accomplish the state preparation, quantum logic gates, and qubit detection. Several important elements of the QCCD architecture have been demonstrated [37-39] that serve as a proof of principle for chip-scale trapped-ion QC, including a recent demonstration of parallel qubit operations in a QCCD device [40].

Another proposed architecture, called a Modular Universal Scalable trapped Ion Quantum Computer (MUSIQC), uses multiple nodes formed by small-scale ion trap devices linked by a photonic quantum interface [41]. Each node is a simple linear ion trap holding a small (10-30) number of qubits. The photonic links allow scaling the system up by adding the nodes and without increasing the complexity of the ion traps themselves.

While both of the above architectures are very promising, it is of great interest to understand if there are other systems capable of storing, cooling, and controlling a large (order of $10^5$) number of ions. The future of QC might rest with distributed systems of moving ions. In contrast to the conventional ion traps, where the ions are stationary and the quantum logic gates are enabled by laser beams focused on individual ions using, for example, multi-channel Acousto-Optic Modulators (AOMs) [42], the ring-like computer will rely on moving ion strings, with ions crossing the interaction region one by one, and quantum gates controlled by a finite set of fixed laser beams, shaped in the time domain by AOMs or the Electro-Optic Modulators (EOMs).

In this paper we attempt to address both of those requirements – i.e. a) the large number of ions, bringing out the notion of a crystalline ion beam instead of a string of few tens of ions and b) the fact that the ions are in motion. We compare a conventional linear ion trap quantum computer setup to a storage ring of crystalline ions, presenting their essential differences, outlining associated challenges, and pointing out directions of future R&D to enable a large-scale QC in a storage ring. In contrast to [30], where the authors in general terms introduced the idea of using storage rings with crystalline beams as quantum computer, we evaluate specific challenges ahead of the ring designers, assess their level of severity and propose mid-term R&D goal to make the first step along long path from the quantum computer on a modern ion trap to the large-scale processor on a future circular machine. We discuss the realistic laser cooling parameters, arrangements and achievable temperatures, propose the specific choice of good qubit candidates (singly-charged ions of calcium and barium), and discuss the quantum gate implementations, and protocols for keeping track of individual qubits in the 100,000 moving ion array.

### 3. General formulation of the problem

For concreteness, we consider an optical qubit in $Ca^+$ and $Ba^+$ ions, which is formed by the ground state ($S_{1/2}$, labelled |g> in the following discussion) and a metastable excited state ($D_{5/2}$ labelled |e>), also called the "shelved state," of the valence electron in these ions. The choice of ions is motivated by their distinctively different atomic mass and optical properties, covering a wide range of ions species suitable for use in QC. The relevant properties of these ions are summarized in Table 1.

| Species | $Ca^+$ [17] | $Ba^+$ [16] |
| --- | --- | --- |
| Isotope | 40 | 138 |
| Lifetime of shelved state | 1 s | 32 s |
| Cooling transition | 397 nm | 493 nm |
| Qubit transition | 792 nm | 1762 nm |
| Typical transverse oscillation frequencies, MHz | 2.5 | 1.0 |
| Typical longitudinal oscillation frequencies, MHz | 1.0 | 0.25 |

Table 1: Comparison between types of ions for QCs

Both ions species have been used in a variety of quantum computation protocols [17, 16], and thus form a useful testbed for our study. Throughout this paper we use a simple model of the ion states, described in Figure 1(A). Here, the yellow arrow represents the qubit transition, corresponding to single-qubit quantum gates, whereas the blue and red arrows are the spectral sidebands, which may be used for sideband cooling and multi-qubit entangling gates. The spectral sidebands are caused by the harmonic motion of ions in the trap whose frequency (referred to as the phonon frequency in Fig. 1, i.e. $|g, 0\rangle$ to $|g, 1\rangle$ transition) is of order $10^6$ Hz, while the qubit transition frequency $|g\rangle$ to $|e\rangle$ is of order $10^{14}$ Hz (~$3.78 \cdot 10^{14}$ Hz in Ca$^+$ or ~$1.70 \cdot 10^{14}$ Hz in Ba$^+$). The machine geometry is shown schematically in Figure 1B.

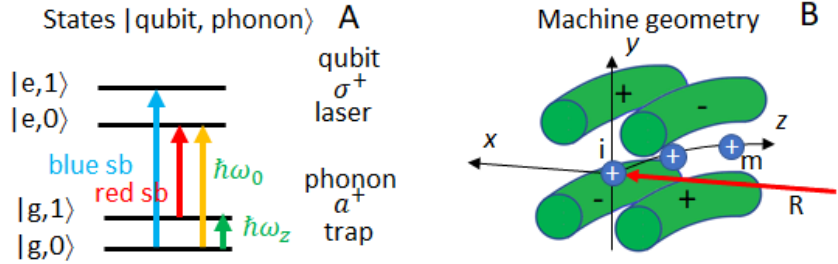

Fig. 1: (A) Simplified diagram of the valence electron state in the ion and indication of the energy levels. The qubit transition (yellow arrow) is between the ground state |g,0> and the excited state |e,0>, while the sideband transitions are shown with red and blue arrows. (B) The machine geometry with the ion chain orbiting the ring.

To illustrate the comparison of the two concepts (a linear ion trap and a storage ring), we write the Hamiltonian governing the state and the dynamics of ions in a crystal revolving in a circular storage ring as a sum of several terms. These terms describe the internal (qubit) state of ions as a two-level system ($H_I$), the dynamics of the particle moving around a circle in bending electric fields ($H_T$), the phonon modes within the ion crystal with N confined ions interacting with each other via Coulomb forces ($H_C$), and, lastly, the atom-laser interaction term for individual ions ($H_L$).

The internal state of individual qubits is that of a two-level system and is equivalent to the state of a spin-½ system, with the Hamiltonian given by

$$H_I = \frac{1}{2}\sigma_3 \hbar \omega_0 + h.c., \qquad (1)$$

which corresponds to the transition between the internal levels in the ion separated by energy $\hbar\omega_0$, with $\sigma_3$ being the z-component of the Pauli matrix responsible for changing the spin state after absorption or emission of a photon with frequency $\omega_0$.

The second Hamiltonian $H_T$ describes the motion of a single ion orbiting in a storage ring:

$$H_T = \frac{p_z^2}{2m} + \frac{x\Delta E}{R} \tag{2}$$

Here, x and z are the components of the position vector, $p_z$ is the component of the momentum vector along the direction of motion, m is the ion mass, $\Delta E$ is the energy deviation from the nominal value, and R is the radius of curvature of the orbit. We note that in the conventional linear ion trap $H_T$ is equal to zero, since the ions are at rest ($p_z \equiv 0$) and form a straight line ($R \equiv \infty$).

The third Hamiltonian $H_C$ captures the interaction of the ions in the crystal via the Coulomb force, which can be decomposed into a spectrum of harmonic oscillations.

$$H_C = \sum_{i=1}^{3N} \hbar \omega_i (a_i^+ a_i + 1/2) \tag{3}$$

Here $\omega_i = \omega_{i,(x,y,z)}$ are the frequencies of the secular motion along the respective axes and $a$ and $a^+$ are the harmonic oscillator annihilation and creation operators, respectively.

The fourth Hamiltonian $H_L$ connects the three interactions, i.e. with the laser field, with the trap field, and with the Coulomb field of the crystal. The resulting expression is obtained by applying the rotating wave and dipole approximations and dropping the higher order terms:

$$H_L = \frac{\hbar \Omega}{2} \sigma^+ exp\{-i\eta(a\, e^{-i\omega_z t} + a^+ e^{i\omega_z t})\} + h.c. \tag{4}$$

Here, $\Omega$ is the Rabi frequency of the transition, $\sigma^+$ is the spin raising operator, and $\eta = \sqrt{\frac{\omega_R}{\omega_z}}$ is the Lamb-Dicke parameter, with $\omega_R = \hbar k_z^2/(2m)$ being the atomic recoil frequency. It is important to note that the nature of the interaction between the laser and the ions in a storage ring is essentially pulsed, since the particles are moving with constant velocity and pass through the fixed laser beam with a finite width.

Having written the principal energies of the problem, we are now equipped with the formalism to look for differences between a typical linear ion trap used in modern QC apparatus and a storage ring. To facilitate this comparison, we selected two subjects for the following discussion. We took an example of the ion trap from the University of Washington linear ion trap experimental set-up [16] and for the storage ring we picked PALLAS from LMU [26].

We emphasize the following differences between a storage ring and a linear ion trap:

- large number of ions,
- ions are in motion,
- motion takes place around a macroscopic circular-like orbit,
- ions are effectively interacting with a pulsed laser beam.

We detail these differences in the subsequent chapters of the paper and evaluate gaps between the main parameters of the ion crystals achieved in rings versus those required for QC. To conclude the paper, we outline some critical technologies that need to be established to bridge this gap and enable a large-scale QC in a storage ring.

### 4. A moving ion

In a storage ring, the ions move around a closed, circle-like orbit with a constant velocity, with an energy that ranges from 1 eV [26] to on the order of 100 keV [43] in mass-spectrometry setups, to sub-TeV in high energy physics colliders [44]. The ions' motion around the machine orbit always comes with some velocity spread due to the finite phase space volume occupied by the beam. The velocity spread corresponds to the finite temperature of the ions' motion. Since the beam consists of many ions, it is customary to adopt a statistical beam description, with the longitudinal temperature averaged across all ions as: $k_B T_{||} = \frac{m \langle \Delta v_z^2 \rangle}{2} = \Delta E$. Dispersion of the particles' motions, expressed in the last term of $H_T$ in (2), defines the ions' temperature due to the difference in the energy of ions orbiting at different radii around the machine. As shown in [33], the apparent $T_{||}$ of the ions' motion is proportional to $1/Q_x^2$, where $Q_x$ is the horizontal tune of the machine.

On the other hand, another limitation of the ions' temperature, this time the transverse one, $T_\perp$, comes from the finite focusing strength of the ring optics, expressed by the third term in $H_T$ in (2). The apparent temperature [33] is proportional to $\mu_{cell}^2$, where $\mu_{cell}$ is the betatron phase advance per cell.

Considerations above lead to the two conclusions on limitations associated with the ions' motion. The first conclusion is that both $T_\perp$ and $T_{||}$ favor rings with high periodicity (or large number of cells) and strong focusing (or high value of the horizontal tune). In this regard, if we compare a synchrotron-like ring with an RFQ ring, we will see a clear advantage of the latter. Indeed, in Table 2 we compared the ASTRID separate functions ring [43] with the PALLAS RFQ [26].

| Device | PALLAS | ASTRID | Ion trap |
| --- | --- | --- | --- |
| Radius between electrodes, mm | 2.5 | N/A | 0.5 |
| Length of ion string, mm | 36.1 | 40000 | <1 |
| Number of ions | 1E4 | $2.8 \cdot 10^5 \ldots 2.8 \cdot 10^8$ | <100 |
| Ions' energy of motion, eV | 1 | $10^5$ | 0 |
| Distance between ions, um | 20 | 15 - 140 | <10 |
| Oscillation frequency $\omega_z$, kHz | 180 | N/A | 200 |
| Oscillation frequency $\omega_{x,y}$, kHz | 110 | 50 | >1000 |
| Horizontal Tune, $Q_x$ | 50 | 2.3 | N/A |
| Periodicity, P | 800 | 4 | N/A |

Table 2: Comparison of two types of storage rings with crystalline beams and a linear ion trap.

The second conclusion is that both of the apparent temperatures, $T_\perp$ and $T_{||}$ are above the ~100 µK range for a realistic ring. Calculating the corresponding apparent temperatures [33], we obtain 1 mK and 0.2 mK for the case of ions in PALLAS, which is consistent with the reported values [31].

In comparison, we estimate the extent of the length of localization of a single ion by taking the second term in $H_T$ in (2), together with $T_{||}$, required for quantum computations in the modern ion traps. We get:

$$\Delta z \approx \sqrt{\frac{2k_B T_{LD}}{m\omega_z^2}} \approx 47 \text{ nm}$$

Here $T_{LD}$ is the temperature corresponding to the Lamb-Dicke regime (approximately 20 µK). For the transverse planes we can use a similar expression [43], where instead of $\Delta z$ we can connect the transverse beam size to the temperature of the corresponding motion $T_\perp$.

Going back to a storage ring like PALLAS, we take the value of $\Delta z$ to be about 7 µm, we get 28 mK for the transverse temperature, which is substantially higher than the value required for a quantum computer employing transverse phonon modes.

The gap between these inherent temperatures of the ring crystals achieved to date and the levels of temperature related to the kinetic energy in a given motion mode of the modern QCs on the ion traps, draws strict requirements for the level of cooling, required to enable a QC, based on a storage ring. There is indeed a very large difference (x1000) in the temperatures of the ion crystals achieved at PALLAS compared to that in the ion traps, which leads to the need for R&D focused on enabling much colder ion beams in rings dedicated to quantum computing experiments.

This value $\Delta z$ is to be compared with a storage ring; for example, with PALLAS, where the crystalline beams move along an orbit that is 36 cm long. Now we will focus on the implications of the macroscopic dimensions of the ring orbit, which, in the case of PALLAS is 36 cm long, whereas, in a typical ion trap, particles are localized within a microscopic length scale on the order of 10-100 microns. There are two effects that arise from the fact that the orbit may not be perfectly aligned with the RF null (the ideal orbit defined by the centers of successive RFQs) on a macroscopic scale: the excess micromotion [45] and the AC Stark and Zeeman shifts caused by the oscillating electric and magnetic fields.

Micromotion is the driven motion of the charged particles in the RF trap at the frequency of the applied RF. Some degree of micromotion is unavoidable and, as long as its amplitude is small enough that the resulting frequency modulation of the laser light used for the qubit logic gates is below 10%, its effect on the quantum gates is negligible. However, any excess micromotion must be minimized to meet this limit.

Excess micromotion occurs when the trapped ions' equilibrium position is shifted away from the RF null by stray electric fields. In a typical linear RF trap setup, where the ion crystal is localized to a small region near the trap center, the stray electric field is uniform (to first order). To compensate for such a field and to bring the ions' equilibrium position close to the RF null, static offset voltages may be applied to the RF electrodes of the trap, or two additional dedicated DC electrodes may be introduced. In a storage ring, however, the ion crystal occupies an extended orbit, and the stray electric field is different at different locations on the orbit. Thus, to compensate for the excess micromotion in a storage ring, a distributed network of compensation electrodes for orbit correction will be required.

Similarly, a distributed system of the beam diagnostics devices will have to be in place to detect the crystal beam orbit in multiple locations around the ring, so that it can be corrected. In the ion storage rings, the crystalline beam diagnostics are realized either with Schottky monitors or using detection of resonant fluorescence. For the low-current beams forming crystalline structures, the resolution limits for Schottky pick-ups become an obstacle, therefore one needs to focus on realizing by a few laser-ion interaction regions where the detectors will be registering the flux of fluorescence photons. Optical detection systems, capable of measuring an ion crystal length, width, momentum spread, and spectra have been developed at ESR, demonstrating good accuracy [46]. Recent progress with development of broadband diode lasers and short-pulse sources <u>motivates future R&D on a compact and highly accurate set-up for diagnostics of ion crystals around macroscopic ring orbit</u>. An accuracy of better than 10 microns in transverse planes is required and, together with the ability of measuring beam size and length, will enable the feedback needed for reaching stable phonon modes in the orbiting ion crystals.

A stray electric field $E_{DC}$ =10 V/m applied perpendicular to the ion orbit in the storage ring trap with the radial secular frequency $\omega_x = 2\pi \times (200\ kHz)$ would shift the equilibrium position of a $^{40}$Ca$^+$ ($^{138}$Ba$^+$) ion by $\Delta x \simeq eE_{DC}/(m\omega_x^2) \simeq 15\ \mu m$ (4.5$\mu m$) [45]. The resulting coherent motion with an amplitude of approximately 0.45 $\mu m$ (0.13 $\mu m$) (assuming the RF frequency of 10 MHz) would carry 16 meV (4.6 meV) of average kinetic energy, with the equivalent temperature of about 190 mK (55 mK). We note, however, that since the micromotion is driven, and not random, this effective temperature does not directly lead to the reduction of the quantum gate fidelity.

The extended orbit of the ions in a storage-ring trap presents a unique challenge to the micromotion compensation. In a typical linear RF trap setup, the ions occupy a very small, essentially one-dimensional region of space on the order of 100 µm long. Any stray DC electric field can be assumed uniform in this case, and a pair of compensation electrodes can be used to bias the trap and minimize the excess micromotion.

The second effect is that of the focusing fields in an RFQ ring on the internal state of an ion when the orbit is somewhat misaligned with respect to the center of the trap. We write the expressions for transverse electric and longitudinal magnetic fields in PALLAS [34] and, taking the misalignment of 10 $\mu m$, obtain that the orbit will sample the oscillating magnetic field of ~20 Gs amplitude and oscillating electric field of 300 V/m amplitude. Both fields oscillate at an RF frequency of 10-100 MHz. These oscillating magnetic and electric fields may cause AC Zeeman and AC Stark effects, respectively, which shift the electronic energy levels of the ions. While the value of the orbit shift will vary along the orbit, each ion will follow the same path (on average) and the relevant figure is the average value of the AC Zeeman and AC Stark shift. For optical qubits with transition frequencies in the 10$^{14}$ Hz range, these fields oscillating at 10-100 MHz would cause negligible effects. However, if hyperfine qubits with ~10 GHz transition frequencies were to be used, the AC Zeeman effect may become noticeable and will need to be taken into account for the qubit transition frequency correction.

In the storage-ring trap, the stray electric field will vary along the orbit, and will have to be compensated by a distributed network of electrodes. In a chip-scale ring trap [47], such DC bias electrodes are located every 15 degrees along the circular orbit. However, we note that precise micromotion compensation (to under 1 µm) is only of essence at the location(s) of the laser beams that drive the quantum gates, while in the remaining part of the orbit only a coarse compensation (on the order of 10s of microns) would be sufficient. Designing an appropriate compensation infrastructure is another important R&D goal.

### 5. Moving ion crystal

Analysis of the Hamiltonian $H_C$ in (3), together with the longitudinal terms from $H_T$ in (2), leads to the discovery of phonon modes for the chain of ions stored in a trap. Indeed, the secular motion of trapped ions in the trap potential can be expanded based on the normal modes of the crystal. Each ion adds three normal modes, so with $10^5$ ions in the crystal, there are three $10^5$ normal modes, each with a different frequency. The frequencies lie in a relatively narrow range, and with so many ions will form an *energy band*.

As the phonon modes are three-dimensional, one could use either longitudinal or transverse ones [48] for creating entanglement between the ions, grouping them into a string of qubits for quantum computation. For the purpose of the current paper we selected the longitudinal modes, which can be described as spectral lines for the collective motion of N coupled oscillators.

Control over the mode frequencies and the coupling strength comes from the ability to confine the ion crystal azimuthally by adjusting the strength of longitudinal focusing. While in the ion traps the longitudinal potential well is realized by supplying DC and AC voltage on the opposite electrode bounding the stationary ion string, in the crystalline storage rings the longitudinal ion density is defined by the number of stored particles.

There have been experiments with an RF bucket at PALLAS [26], where the AC voltage, synchronous with the beam motion, was applied to electrodes between RFQ segments to generate a potential well. The goal of the experiment was to develop techniques of manipulation with the linear ion density to generate various crystalline structures.

The experiment was partially successful, in the sense that the ions clearly exhibited bunching, following the potential well pattern. However, the linearity of the ion string was only preserved at low bunching voltages and, as soon as the voltage moderately increased, the structure of the ion crystal morphed from the chain to a zigzag. Also, due to the parabolic shape of the bunch, the ions' spacing was not linear, leading to the zigzag phase originating in the middle of the bucket, while the edges of the crystal were still chain-like. Applying this method, i.e. RF modulation on one of the chambers of an RFQ ring, to the case of the quantum computer on a storage ring, we see a challenging complex spread in phonon frequencies being affected by the nonlinear shape of the potential well. Thus, one of the serious issues, which is yet to be solved for future storage rings for QC applications, is the bunching or grouping techniques for the ion crystals.

The operation of quantum gates is based on the requirement that ions stay in the motional ground state. In the case of the fast-moving ions, their transport inherently creates a motional excitation of their state. The challenge is to develop transport protocols that minimize energy transfer between each ion's motion and the phonon coupling between ions. It has been demonstrated [36-39] that non-adiabatic shuttling operations preserve the final state of the ion close to the motional ground state. It has also been shown that quantum information stored in both the motional and the spin degree of freedom is preserved during the process of shuttling.

In particular, in experiments [38], a single $^9$Be$^+$ ion was transported by 370 μm in 8 μs, with an energy gain of only 0.1 motional quanta. Similar results were achieved for the transport of two ions, as well as for separating chains of up to 9 ions from one to two different potential wells. Separating two ions was accomplished in 55 μs, with excitations of about two motional quanta for each ion.

These encouraging results are still very far from the specifications necessary to enable a quantum computer with moving ions in a storage ring. Indeed, preserving the integrity of the phonon modes at the level of single motional quantum appears to be a tough goal, when one considers a meter-long orbit circumference where the particles are affected by DC and AC fields. <u>Designing such low-noise machines is one of the R&D efforts needed to demonstrate the suitability of the rings for quantum computations</u>.

### 6. *Moving ion crystal interacting with a laser*

We consider the Doppler cooling first. To efficiently Doppler-cool the trapped ion crystal, the red-detuned laser beams must have significant wave vector components along all three principal axes of the ion trap. In a storage ring, there is an additional consideration of the lack of the ion confinement along the orbit. If not accounted for, this would lead to the ion crystal being accelerated by the radiation pressure of the Doppler cooling laser. This effect, however, presents a unique opportunity for the ion crystal rotational velocity control.

Let us consider the Doppler cooling laser arrangement shown in Figure 2. The two counter-propagating beams are both red-detuned from the atomic resonance. If the detunings are identical, and both beams have the same intensity, then the radiation pressure from both beams on the ions are equal and the ions will be stationary. However, if the frequencies of the two laser beams differ by amount $\Delta\omega$, then the ions will move in the direction of the beam that has a smaller lab frame detuning until, in the rest frame of the ions, both lasers have the same detuning due to the Doppler effect. This condition is satisfied when the velocity of the ions is given by $kv = \Delta\omega/2$. For example, if the desired Ca$^+$ ion velocity is 100 m/s, then the detuning between the two 397 nm Doppler cooling lasers needs to be approximately 80 MHz.

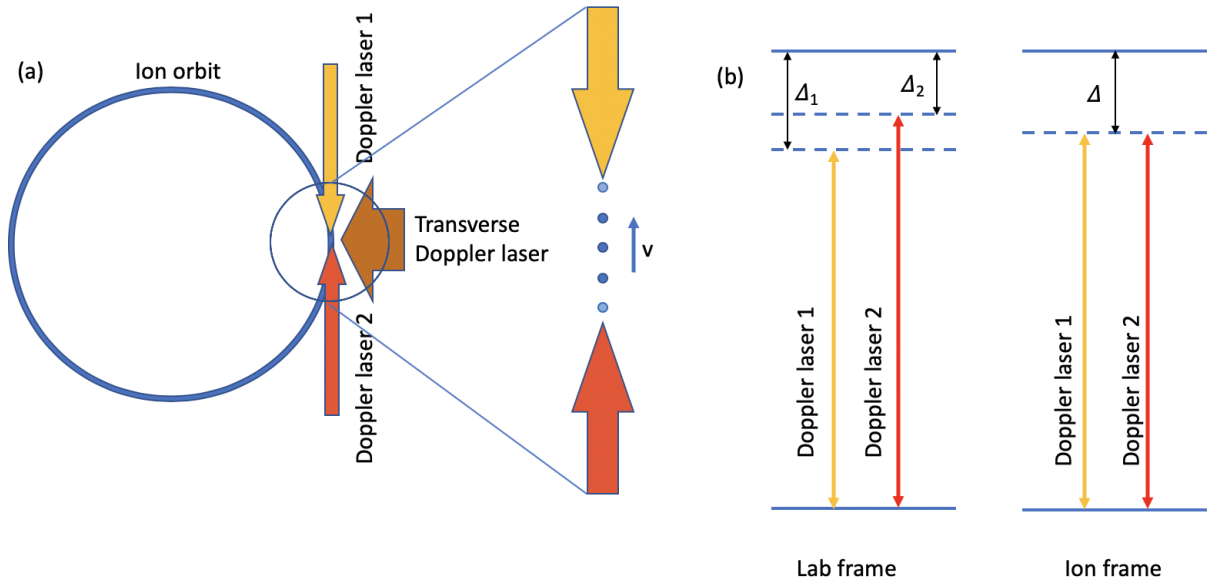

Fig. 2. Doppler laser cooling beam arrangement. (a) Two focused, counter propagating laser beams ("Doppler laser 1" and "Doppler laser 2") with different frequencies but equal intensities are applied tangentially to the ion orbit to cool the ion motion in the z-direction, as well as to set the orbital velocity. The third beam ("Transverse Doppler laser"), shaped to have an elongated elliptical profile along the ion orbit direction, is applied perpendicular to the orbit to enable cooling of the x and y directions. (b) The relevant atomic energy levels and laser detunings in the Lab frame (left) and the moving Ion frame (right). In the Lab frame, the detuning $\Delta_1$ of the "Doppler laser 1" is greater than the detuning $\Delta_2$ of the "Doppler laser 2." This creates a larger scattering force by the "Doppler laser 2," so that the ions are forced to move in the direction of that laser until their velocity is such that the detunings of both lasers are the same ($\Delta$) in the Ion frame. Then the scattering forces due to both lasers are equal, and the ions move with a constant speed.

The counter-propagating Doppler cooling beam arrangement does not, however, efficiently cool the transverse motion of the ions. To address this problem, a third laser beam may be used that is perpendicular to the orbit. For more efficient interaction with the ions, the laser beam's transverse cross section may be made elliptical, extended along the orbit, using cylindrical lenses.

*Quantum gates*

Now we look at the implications of the laser interacting with ions as described by $H_L$ in (4). Quantum logic gates on trapped ions are typically driven by the laser beams whose intensity, frequency, and phase is controlled by an RF signal generator via an AOM. After passing through the AOM, the laser beams are focused onto the ions in the trap. Following this model, we consider a laser beam, interacting with the chain of ions, which are moving through the interaction region, as opposed to being static in a conventional ion trap. We assume that the crystal orbits around the ring at $v$ = 2.8 km/s (1 eV of kinetic energy for $^{24}$Mg$^+$) as it was actually observed in the PALLAS RFQ ring. A slower beam would relax the requirements on the gate speeds that we find below.

We estimate the separation of ions moving around with velocity *v* in a crystalline beam to be approximately 20 μm ( $\Delta \approx \sqrt[3]{\frac{e^2}{2\pi\varepsilon_0 m\omega_z^2}}$, [33]). The laser beam crosses the ions' path with a waist of $w_0$ = 10 μm, which corresponds to half of the separation between consecutive ions in the chain, so that the neighboring ions are not affected by the laser beam controlling the state of a given qubit. This gives the Rayleigh range of the laser beam $z_R = \frac{\pi w_0^2}{\lambda_l} \approx$ 0.8 mm. The laser beam envelope as a function of the distance transverse to the orbit and the chain of ions (blue circles), together with the laser intensity distribution at the focus of the laser beam, are shown in Figure 3.

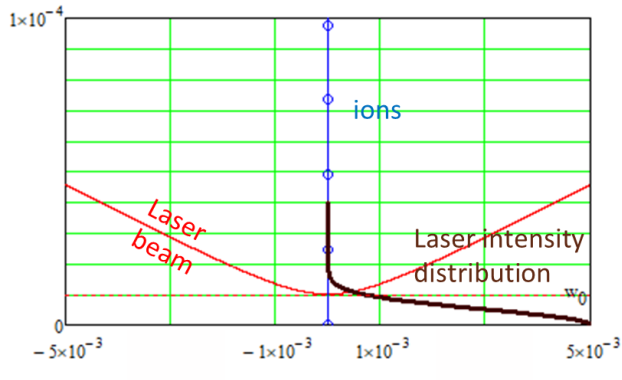

Fig. 3: Model of the Quantum Gate, illustrating the laser Gaussian beam transverse envelope in red, intensity distribution of the laser beam at the waist in brown, and positions of ions in the crystalline beam (blue), which is moving from bottom to top and traversing the laser beam in the interaction region. The laser waist is a half of the distance between the consecutive ions and the laser pulse length is a half of the time of flight of the ions through the same region. A single laser pulse interacts with a single ion setting the qubit's state prior to execution of a cycle of quantum computation.

The laser source should be on during the time of flight of an ion through the interaction region and then the light should be turned off so as not to affect the next ion in the chain. Turning the laser on and off with a high-bandwidth AOM can be done with ~2 ns rise/fall times. The pulse length is then $\tau_l = \frac{w_0}{2v} \approx$ 4.6 ns, where a factor of 2 is taken to accommodate the ramp-up of intensity along the pulse. This is significantly shorter than the typical trapped ion quantum gate times (approximately 1 μs for single qubit gates and 100 μs for two-qubit gates when CW laser beams are used). We note that quantum gates as fast as 3 ns have been reported using mode-locked lasers [49].

The laser intensity necessary to drive the gates in this limited amount of time is defined by the Rabi oscillation frequency for the qubit transition. For example, a bit-flip operation on a single qubit is equivalent to a Rabi π-pulse, such that the probability of the qubit transition from state |g,0> to state |e,0> (and vice versa) is $P = \sin(\Omega\tau_l/2)^2 = 1$. For the qubit transition in Ca$^+$ at 792 nm (Ba$^+$ at 1762 nm), which is an electric quadrupole transition, the Rabi frequency at the laser intensity of ~5000 W/mm$^2$ (250 mW/mm$^2$) was measured to be approximately $2\pi \cdot$1000 kHz ($2\pi \cdot$43 kHz) [17, 50], which gives a π-pulse time of about 0.5 μs (11.6 μs). For an electric quadrupole transition, the Rabi frequency scales linearly with the laser intensity; thus, to achieve a 4.6 ns $\pi$-pulse time, a laser intensity of about 540000 W/mm$^2$ would be needed for Ca$^+$ (about 600 W/mm$^2$ would be needed for Ba$^+$). This is a manageable level of intensity for barium and can be realized with a 100 mW CW laser focused to a 10 $\mu m$ waist. The requirement for Ca$^+$ is much more severe. It may be possible to enable step-wise gates executed over many cycles of rotation that would relax these power requirements.

For two-qubit entangling gates, the standard protocol in the trapped ion QC is to use the quantized collective motion of all ions in the crystal as the quantum "data bus" to couple the individual qubits [51, 52]. Such gates have to be performed by adiabatically exciting the quantized ion motion using lasers that couple the qubit spin and motional states, as in the Hamiltonian presented in (4), in the resolved sideband regime. However, with ~10$^5$ normal modes achieving this regime becomes impractical. For such a large and distributed set of qubits, the all-to-all qubit connectivity, which is enabled by the resolved sideband regime in a linear ion trap, cannot be realized, and different types of two-qubit and multi-qubit gates will have to be developed that involve only the nearest neighbor qubits [51-53]. Optical power requirements for these gates to be implemented on the ns time scales is likely to significantly exceed the power available from the existing CW laser sources. <u>Development of the novel protocols that involve ultrafast pulsed lasers is another direction for the R&D</u>. We note a recently proposed "linear-tape" architecture of a trapped-ion quantum computer [56], which bears some relevance to the storage-ring architecture and may serve as a useful starting point for future development.

To realize such fast quantum gates, an adequate AOM system that is capable of addressing every ion in the crystal is necessary. This will require the AOM to switch at the rate of 140 MHz while generating short pulses with ~2 ns rise/fall times and tunable intensity and polarization. Such fast modulators are available commercially (for example, the GPM-1600-400 AOM by Brimrose [42]) but may not be able to handle the amount of optical power needed. We thus conclude that future quantum computers on crystalline beams will rely on broadband modulators that will be capable of handling high optical power while generating short pulses with high RF bandwidth in the range of 100s of MHz. <u>This is another important direction for future R&D focusing on realizing quantum computing with moving ions.</u>

One other important conclusion is that the ions' velocity plays a pivotal role in every aspect of the pulsed and time-of-flight techniques that will be used in quantum computers. The minimal sustainable value of the ion's velocity is limited by the balance of ion heating by various environmental factors and field errors, and the efficiency of cooling being localized to a fraction of the ring circumference. <u>Reducing the ions' velocity in storage rings while preserving the strong focusing of ion orbits and a high degree of cooling is another formidable task for future R&D.</u>

### 7. *Large number of ions*

In addition to the laser cooling and orbit control problems, the very large number of ions in the storage-ring quantum computer presents other unique challenges relevant to quantum computing operations: (1) keeping track of the individual qubits; (2) identifying and mitigating qubit loss due to collisions with the residual background gas; and (3) detecting ion reordering due to such collisions.

(1) We propose a stochastic method for keeping track of the individual ion qubits. When loading the ring, another ion species is randomly mixed with the main qubit ion in a small proportion (1:10 ratio or so). The other species may be simply a different isotope of the same species as the main qubit, such as $^{43}$Ca if the main qubit ion is $^{40}$Ca or $^{136}$Ba if the main qubit is $^{138}$Ba. Then, when laser-cooled, the admixture ions will not fluoresce and will present themselves as dark gaps in the chain of bright qubit ions. These bright/dark patterns will be random and unique at every point along the linear ion crystal, thus allowing one to identify the qubit ions. Figure 4 shows an example of a small trapped barium ion chain with the $^{138}$Ba ions being bright and ions of a different isotope ($^{136}$Ba) being dark.

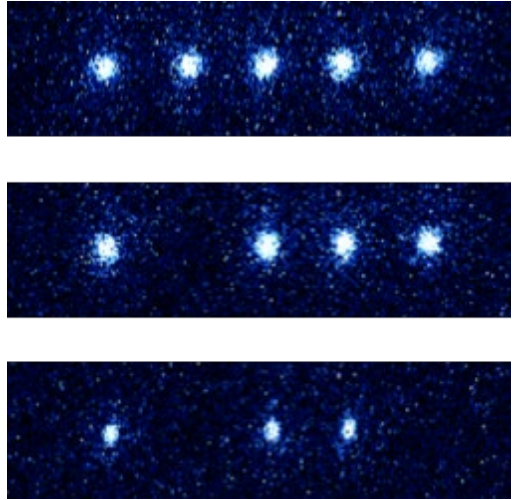

Figure 4. Intensified CCD camera images of trapped Ba ions in a UW linear trap. Only the $^{138}$Ba ions are resonant with the Doppler cooling laser, so only they are visible. $^{136}$Ba ions appear as dark gaps in the chain. Top: A five-ion chain composed of five $^{138}$Ba ions. Middle: A five-ion chain composed of four $^{138}$Ba ions and one $^{136}$Ba ion (second from the left). Bottom: A five-ion chain composed of three $^{138}$Ba ions and two $^{136}$Ba ions (second and fifth from the left).

(2) and (3) If a mismatch in the bright/dark pattern is detected, it may be due to ion loss or ion reordering as a result of background gas collisions. If such an event is detected, then the subset of qubits involved in it must be reset and re-encoded. <u>Development of fault-tolerant protocols that allow mitigating such problems is another direction for future R&D.</u>

The "switching rate" of an ordinary trapped ion quantum processor is limited by the minimum duration of the laser pulses interacting with the ions necessary to complete a quantum gate. According to Steane [11], an estimate for the maximum Rabi frequency to drive motional sidebands without affecting the nearby quantum motional states in a crystal with N ions is given by $\Omega_N = \eta\Omega/\sqrt{N}$, where the Rabi frequency $\Omega$ corresponds to a single ion transition. Then the switching rate is set by the inverse of the Rabi $2\pi$-pulse time.

Two problems arise here when considering a storage-ring QC architecture. The first is related to the duration of the pulse, limited by the time of flight of the ion through the ion-laser interaction zone, as described in the previous section. The other issue is scaling the switching rate with the number of ions at large values of N. Indeed, taking the single-ion Rabi frequency of $2\pi \times 218$ MHz, as required for a single qubit gate to be completed in 4.6 ns, we estimate the gate time for 100 ions as 230 ns (taking $\eta = 0.2$). For $10^5$ ions the gate time becomes 7.3 μs, which severely limits the computing rate.

Another, more severe limitation on the gate speed comes from the fact that the phonon spectrum of a crystal of $10^5$ ions is very dense. If the transverse center of mass frequency is of order 1 MHz, then the $10^5$ normal modes lie in a band of approximately 1 MHz in widths, thus leaving approximately 10 Hz per mode. Thus, in order to achieve the resolved sideband regime, the operations must be performed slower than 0.1 s. We conclude that the regular approach used in the ion traps for performing quantum gates is not suitable for the storage ring QC, and fast quantum gates will need to be developed that do not use a single normal mode addressing [53-55]. Piecewise gate implementation may also be of interest here, as mentioned above, where the entire gate is split into a series of smaller steps, each completed during one passage of the ions through the interaction region.

## 8. Summary

In this paper we analyzed a new, interesting concept: turning storage rings with crystalline beams into a giant trap for quantum computing. It is clear that there are gaps between the quality of crystalline beams achieved to date and what is required to enable quantum computations. The motion of particles along a macroscopic orbit, the high thermal temperature of the ions' motion, the limited time for interaction with cooling and gate lasers, sensing and correcting the ions' motion – these and other challenges present clear objectives for future R&D.

Below we focus on each of the seven R&D directions that we outlined in this paper. Following the flow of out logic above, we move from a single ion to the moving ion crystal and, then, to that with a large number of ions, interacting with lasers.

The first challenge is to produce much colder beams than those available today and learn how to sustain their low temperature in steady state. The goal of reducing the temperature from the 1 mK range, demonstrated at PALLAS, to 0.1 mK would be the first significant step towards producing beam quality suitable for enabling quantum computations. Proof-of-principle experiment, based on the Doppler cooling followed by the EIT cooling [21], is conceivable to, first, model and, then, design for a specialized storage ring.

The second challenge is to advance engineering of storage rings to reduce the amount of thermal and RF noise. An interesting experiment could be to include into an existing crystalline ion ring a straight section, such at IOTA, with broadly variable temperature and, using narrowband laser as a "thermometer", study the ion beam heating rates and lowest achievable temperatures.

The third challenge is to develop a compact and highly accurate diagnostics and control systems of the orbit of ion crystals around the macroscopic ring. Design of a single-particle beam position monitor (BPM), either photon or RF, with the goal of demonstrating both the turn-by-turn detection and slow acquisition, of the single particle and crystalline beam position with an accuracy of 10 um appears to be an exciting goal for future engineering of QC beam diagnostics. We foresee using high-fidelity detection of the laser-induced fluorescence for the photon BPMs and cryogenically cooled 4-button set-up for the RF BPM similar to those used in the Penning trap [57]. The latter, if feasible, can also be purposed for the orbit correction system by integrating DC-offsets to the sensing button or wires.

Interaction with lasers present a number of specific and principal challenges. The fourth challenge is to develop novel protocols for two-qubit and multi-qubit gates that involve ultrafast pulsed lasers. The goal here is to build theoretical models for the basic quantum logic operations suitable for the large, moving ion crystals. Experimental demonstrations of the feasibility of such quantum gates may be performed in the regular linear ion traps where the motion of the ion crystal can be induced by modulating the endcap electrode voltages.

As we have shown above, the ultrafast lasers appear to be necessary to enable these quantum logic gates. The fifth challenge is to design and build a broadband modulator that is capable of handling high optical power of such lasers while generating short pulses with high RF bandwidth in the range of 100s of MHz. A multichannel AOM producing 2 ns rise and fall time pulses at a gigahertz rate, while sustaining Watts of optical power with intensities of order $MW/mm^2$ stands as a goal for realization of the quantum logic gates in the moving ion crystals.

On the other hand, we saw that the straightforward method of alleviating tight requirements on the laser systems and AOMs is reducing the ions' velocity in storage rings while preserving the strong focusing of ion orbits and a high degree of cooling. Addressing this, sixth, challenge, we arrive at the need of developing a "thermodynamic" particle tracking code that takes into account radiation, absorption of photons, finite temperature of surrounding chambers, and vacuum conditions. The code should also be capable of computing noise in the particle beams, cooling effects, evaluating and keeping track of the quantum state of particles and their Coulomb fields during beam propagation around the ring. Equipped with such code, we will be in the position to make headway in optimizing the ring models under conditions of heating to test the use of complex cooling techniques and to understand the limit of lowest ion crystal velocities.

Protecting the quantum states of qubits from decoherence and loss of quantum states in a storage ring setup requires development of specific fault-tolerant protocols. While standard quantum error correction protocols used in ion trap may be applicable to combat the decoherence, there are special considerations for large moving ion crystals. The seventh goal is to develop theoretical models that enable: (a) keeping track of the individual physical qubits as they move around the orbit; (b) identifying and mitigating qubit loss due to collisions with the residual background gas; and (c) detecting ion reordering due to such collisions. Numerical simulations may be used to verify the effectiveness of these protocols initially, followed by experimental demonstrations in linear ion trap.

Concluding our paper and being now equipped with considerations on the scale of the problem, we may ask ourselves: how conceivable is the concept of macroscopic storage rings running $10^5$ ions, entangled and coupled to the RF pick-ups and lasers? At this point it is difficult to say, since the gaps between what has been demonstrated and what is required are so large. However, the reward for putting such a system to work is dramatic: the highly redundant and reliable operation of the fastest-ever computer with parallel processors, capable of reaching new horizons in image recognition, data mining, and solving other problems of extreme complexity. We believe that when the seven objectives described above were met, we will be in the position to design an experiment and to code the very first simple quantum program in a crystalline beam of particles revolving around a macroscopic storage ring.